

\input vanilla.sty 




\mag=\magstep1
\baselineskip=20pt
\hsize=16truecm
\vsize=23truecm
\pageno=1
\TagsOnRight

\font\tenbf=cmbx10
\font\tenrm=cmr10
\font\tenit=cmti10

\font\sevenrm=cmr7

\outer\def\rmproclaim#1{\medbreak\noindent\smc\ignorespaces
    #1\unskip.\enspace\ignorespaces\rm}
\outer\def\endrmproclaim{\par\ifdim\lastskip<\medskipamount\removelastskip
  \penalty 55 \fi\medskip}

\def\=def{\; \mathop{=}_{\text{\rm def}} \;}
\def\del{\partial}
\def\res{\; \mathop{\text{\rm res}} \;}

\def\R{{\bold R}}
\def\Z{{\bold Z}}

\def\L{{\cal L}}
\def\M{{\cal M}}
\def\B{{\cal B}}

\def\zbar{{\bar{z}}}
\def\Bbar{{\bar{\B}}}

\def\Lhat{\hat{\L}}
\def\Mhat{\hat{\M}}
\def\Bhat{\hat{\B}}
\def\Fhat{\hat{F}}
\def\Shat{\hat{S}}
\def\chat{\hat{c}}
\def\uhat{\hat{u}}
\def\vhat{\hat{v}}
\def\zhat{\hat{z}}

\line{Research Institute for Mathematical Sciences \hfill RIMS-790}
\line{Kyoto University August 1991}

\title
    SDIFF(2) TODA EQUATION \\
    -- HIERARCHY, TAU FUNCTION AND SYMMETRIES
\endtitle
\author
    Kanehisa Takasaki \\
    {\rm Institute of Mathematics, Yoshida College, Kyoto University}\\
    {\it Yoshida-Nihonmatsu-cho, Sakyo-ku, Kyoto 606, Japan}\\
    {\rm and}\\
    Takashi Takebe \\
    {\rm Department of Mathematics, Faculty of Science,
         University of Tokyo}\\
    {\it Hongo, Bunkyo-ku, Tokyo 113, Japan}\\
\endauthor
\vglue 2truecm

\vglue 1truecm

\noindent{\bf Abstract.}
A continuum limit of the Toda lattice field theory, called the
SDiff(2) Toda equation, is shown to have a Lax formalism and an infinite
hierarchy of higher flows. The Lax formalism is very similar to the
case of the self-dual vacuum Einstein equation and its hyper-K\"ahler
version, however now based upon a symplectic structure on a cylinder
$S^1 \times \R$.  An analogue of the Toda lattice tau function
is introduced.  The existence of hidden SDiff(2) symmetries are derived
from a Riemann-Hilbert problem in the SDiff(2) group.  Symmetries of
the tau function turn out to have commutator anomalies, hence give
a representation of a central extension of the SDiff(2) algebra.

\newpage
\heading
    1. Introduction
\endheading

\noindent
Two dimensional Toda fields on an infinite chain, $\phi_i, \ i\in \Z$,
are described by the nonlinear field equation
$$
    \del_z \del_\zbar \phi_i
    +\exp(\phi_{i+1}-\phi_i) -\exp(\phi_i-\phi_{i-1}) = 0      \tag 1
$$
or, equivalently, for $\varphi_i=\phi_i-\phi_{i-1}$ by
$$
    \del_z \del_\zbar \varphi_i +\exp \varphi_{i+1}
    +\exp \varphi_{i-1} -2 \exp \varphi_i  = 0.             \tag 2
$$
In continuum limit as lattice spacing tends to 0, $\phi_i$ and
$\varphi_i$ will become three dimensional fields
$\phi=\phi(z,\zbar,s)$ and $\varphi=\del\phi/\del s$ with
an extra space variable $s \in \R$.  The equation of motions are
then given by
$$
    \del_z \del_\zbar \phi +\del_s \exp \del_s \phi = 0     \tag 3
$$
and
$$
    \del_z \del_\zbar \varphi +\del_s^2 \exp \varphi= 0.    \tag 4
$$
{}From this observation, Saveliev and Vershik [1] introduced the notion
of Cartan operators and corresponding (continual) Lie algebras, and
applied them to solving the above three dimensional equations.
Bakas [2] and Park [3] gave another interpretation of these equations
in the language of extended conformal symmetries (area-preserving
diffeomorphisms, $w_\infty$ algebras, etc.).  Since the group of
area-preserving (i.e., symplectic) diffeomorphisms, SDiff(2), plays a
key role in the latter point of view, let us call the equation of
motion of the $\phi$ and $\varphi$ fields the SDiff(2) Toda equation.
The original Toda equation, in this sense, may be called the GL($\infty$)
Toda equation.

In fact, there are further two sources of the SDiff(2) Toda equation.
One is discovered by relativists, Boyer and Finley [4] and
Gegenberg and Das [5], in the study of ${\cal H}$-spaces (heavens)
with a rotational Killing symmetry. Another source is Einstein-Weyl
geometry studied by twistor people [6] [7] [8] [9]
in the context of curved minitwistor spaces.

The contents of this letter are influenced by all these
approaches, but mostly based upon the method of Kyoto group
[10] [11] originally developed for soliton equations.  A direct
prototype is the theory of the Toda lattice (TL) hierarchy
[12] [13] that was proposed as a Toda lattice version of
the KP hierarchy.  Our aim is to give a similar framework
to study the SDiff(2) Toda equation.

We first show a Lax formalism along with a set of higher commuting
flows (hierarchy), secondly, introduce an analogue of the tau functions
of the TL hierarchy, and finally, with the aid of a Riemann-Hilbert problem
in the SDiff(2) group, construct SDiff(2) symmetries of the hierarchy
explicitly.  A remarkable result is that these symmetries exhibit
commutator anomalies at the level of the tau function, hence the
true symmetry algebra should be a central extension of the original
SDiff(2) algebra.  Apart from the treatment of the tau function,
technical details are mostly the same as in the case of the self-dual
vacuum Einstein equation [14]. Unlike that case, however, the spectral
parameter $\lambda$ now plays the role of a true variable: $\lambda$ and
$s$ form a coordinate system on a cylinder $S^1 \times \R$ on which
we consider a symplectic structure and associated symplectic
diffeomorphisms.

\heading
    2. Lax formalism -- heuristic consideration
\endheading

\noindent
A ``zero-curvature representation" of the SDiff(2) Toda equation is
given by
$$
    \del_\zbar \B - \del_z \Bbar + \{ \B, \Bbar \} = 0,      \tag 5
$$
where $\{\quad,\quad\}$ stands for the Poisson bracket
$$
    \{ F, G \}
      \=def \lambda(\del_\lambda F)(\del_s G)
           -\lambda(\del_\lambda G)(\del_s F)                \tag 6
$$
and $\B$ and $\Bbar$ are given by
$$
\align
      \B &= \lambda\exp[(-\alpha +\frac{1}{2})\del_s\phi]
         +(\alpha+\frac{1}{2})\del_z\phi,                             \\
   \Bbar &= \lambda^{-1}\exp[(\alpha+\frac{1}{2})\del_s\phi]
         +(\alpha-\frac{1}{2})\del_\zbar\phi.                \tag 7   \\
\endalign
$$
Here $\alpha$ is a gauge parameter that also arises in the TL hierarchy
[13]; we mostly consider the $\alpha=1/2$ gauge.  The Poisson bracket is
obviously related with area-preserving diffeomorphisms on a cylinder.
This zero-curvature equation is already proposed by Kashaev et al. [15].

We note that similar zero-curvature representations take place in
the case of the self-dual vacuum Einstein equation [14].
The only difference is that $\lambda$ is no more a parameter
(spectral parameter) but a true variable.  Bearing this difference
in mind, we now modify the framework developed therein so as to fit
into the present setting.  We first introduce an exterior differential
2-form as
$$
    \omega \=def \frac{d\lambda}{\lambda}\wedge ds
                +d\B \wedge dz + d\Bbar \wedge d\zbar,     \tag 8
$$
where $d$ here (and from now on) stands for total differentiation
with respect to the space-time variables $z,\zbar,s$ and the spectral
variable $\lambda$. This 2-form is obviously a closed form,
$$
    d\omega = 0,                                           \tag 9
$$
and satisfies the algebraic relation
$$
    \omega \wedge \omega = 0.                              \tag 10
$$
The latter is equivalent to zero-curvature equation (5).  These two
relations imply the existence of two functions $P$ and $Q$ that
give a pair of ``Darboux coordinates" as
$$
    \omega = \frac{dP}{P} \wedge dQ.                       \tag 11
$$
(Note that the Poisson bracket $\{\quad,\quad\}$ is twisted by
an extra factor $\lambda$. The presence of the denominator $P$ is
due to this fact.) From (11) one can deduce the Lax equations
$$
\align
    & \del_z P = \{\B,P\}, \qquad
      \del_\zbar P = \{\Bbar,P\},                   \\
    & \del_z Q = \{\B,Q\}, \qquad
      \del_\zbar Q = \{\Bbar,Q\},         \tag 12   \\
\endalign
$$
and the canonical Poisson relation
$$
    \{P,Q\} = P.                          \tag 13
$$
Conversely, from (12) and (13) one can go back to (11); they are
equivalent. In the case of the self-dual vacuum Einstein equations
[14], Lax equations like (12) are also interpreted as a ``linear
system." This double nature seems to be a common characteristic of
this kind of nonlinear systems (nonlinear graviton equations).

In a generic case (see Section 4), one can find two distinct pairs of
Darboux coordinates, $(\L,\M)$ and $(\Lhat,\Mhat)$, with the following
analyticity properties on the complex $\lambda$ plane.

\item{i)} these four functions are holomorphic functions of $\lambda$
in a neighborhood of a circle $\Gamma$ with center at the origin
$\lambda=0$.
\item{ii)} $\L$ and $\M$ can be analytically extended outside $\Gamma$
up to the point $\lambda=\infty$ where they have first order poles.
\item{iii)} $\Lhat$ and $\Mhat$ can be analytically extended inside
$\Gamma$.

\noindent More specifically, one may select them to have Laurent
expansion as
$$
\align
     \L  = \lambda + \sum_{n\le 0} u_n\lambda^n,   \quad &
     \M  = z\L + s + \sum_{n \le -1} v_n \L^n,     \\
   \Lhat = \sum_{n \ge 1} \uhat_n \lambda^n,       \quad &
   \Mhat = -\zbar\Lhat^{-1}
            + s + \sum_{n \ge 1} \vhat_n\Lhat^n.    \tag 14 \\
\endalign
$$
Note that $\M$ and $\Mhat$ are Laurent expanded in $\L$ and $\Lhat$
rather than in $\lambda$; coefficients thus defined will be used
for defining a tau function. A few Laurent coefficients are
directly related with $\phi$ as
$$
\align
    \del_z \phi     &= u_0,                            \\
    \del_\zbar \phi &= \uhat_2 \uhat_1{}^{-2},         \\
    \del_s \phi     &= -\log\uhat_1            \tag 15 \\
\endalign
$$
One can thus single out two particular pairs of Darboux coordinates
$(\L,\M)$ and $(\Lhat,\Mhat)$ for the 2-form $\omega$,
$$
    \omega = \frac{d\L}{\L} \wedge d\M
           = \frac{d\Lhat}{\Lhat} \wedge d\Mhat.      \tag 16
$$
Of course they both can also be characterized by a Lax system and a
canonical Poisson relation like (12) and (13).

\heading
    3. Hierarchy and tau function
\endheading

\noindent
The notion of the SDiff(2) Toda hierarchy is a straightforward
extension of the above Lax formalism.  We now start from two
pairs of Laurent series $(\L,\M)$ and $(\Lhat,\Mhat)$ of the form
$$
\align
     \L  \=def \lambda + \sum_{n\le 0} u_n\lambda^n, \quad &
     \M  \=def \sum_{n\ge 1}nz_n\L^n + s + \sum_{n \le -1} v_n \L^n,  \\
   \Lhat \=def \sum_{n\ge 1} \uhat_n \lambda^n,      \quad &
   \Mhat \=def -\sum_{n\ge 1}n\zhat_n\Lhat^{-n}
                + s + \sum_{n \ge 1} \vhat_n\Lhat^n,           \tag 17 \\
\endalign
$$
where $z_n$ and $\zhat_n$, $n=1,2,\ldots$, now supply an infinite
number of independent variables along with $s$ and $\lambda$.
The hierarchy consists of the Lax equations
$$
    \del_{z_n}K     = \{ \B_n, K \},  \quad
    \del_{\zbar_n}K = \{ \Bhat_n, K\}                  \tag 18
$$
for $K=\L,\M,\Lhat,\Mhat$ and the canonical Poisson relations
$$
    \{ \L, \M \} = \L, \quad
    \{ \Lhat, \Mhat \} = \Lhat.                          \tag 19
$$
Here $\B_n$ $\Bhat_n$ are given by
$$
    \B_n    \=def (\L^n)_{\ge 0}, \quad
    \Bhat_n \=def (\Lhat^{-n})_{\le -1},                    \tag 20
$$
where $(\quad)_{\cdots}$ stands for extracting powers of $\lambda$
with exponents running over the range shown in the suffix.
These equations can be recast into a compact form as
$$
    \omega = \frac{d\L}{\L} \wedge d\M
           = \frac{d\Lhat}{\Lhat} \wedge d\Mhat,         \tag 21
$$
where the 2-form $\omega$ is now given by
$$
    \omega \=def \frac{d\lambda}{\lambda} \wedge ds
                +\sum_{n\ge 1} d\B_n \wedge dz_n
                +\sum_{n\ge 1} d\Bhat_n \wedge d\zhat_n, \tag 22
$$
and, accordingly, satisfies the equations
$$
    d\omega = 0,  \quad  \omega \wedge \omega = 0.      \tag 23
$$
The second relation of (23) is equivalent to the zero-curvature
equations
$$
\align
    & \del_{z_n}\B_m -\del_{z_m}\B_n +\{ \B_m, \B_n \} = 0,      \\
    & \del_{\zhat_n}\Bhat_m -\del_{\zhat_m}\Bhat_n
                               +\{ \Bhat_m, \Bhat_n \} = 0,      \\
    & \del_{\zhat_n}\B_m -\del_{z_m}\Bhat_n
                               +\{ \B_m, \Bhat_n \} = 0. \tag 24 \\
\endalign
$$
The original SDiff(2) Toda equation is included in the $(z_1,\zhat_1)$
sector with $z_1=z$, $\zbar=\zhat_1$, $\B=\B_1$ and $\Bbar=\Bhat_1$.
We note that Lax equations like (18) containing a Poisson
bracket are already studied to some extent by Golenisheva-Kutuzova and
Reiman [16] by means of the coadjoint orbit method.

We now show that four important potentials are hidden in the above
equations.  The first one is the $\phi$ field itself.  This is
characterized as
$$
    d\phi = \sum_{n\ge 1}\res (\L^n d\log\lambda)dz_n
           -\sum_{n\ge 1}\res (\Lhat^{-n} d\log\lambda)d\zhat_n
           -\log \uhat_1 ds,                              \tag 25
$$
where ``$\res$" denotes the formal residue operator
$$
    \res \sum a_n \lambda^n d\lambda \=def a_{-1}.             \tag 26
$$
The construction of the second and third potentials are inspired by
Krichever's work [17] on a SDiff(2) version of the KP hierarchy.
These potentials, $S$ and $\Shat$, are defined as
$$
\align
    & dS = \M d\log\L + \log\lambda ds
          +\sum_{n\ge 1} \B_n dz_n +\sum_{n\ge 1} \Bhat_n d\zhat_n,
                                                           \tag 27 \\
    & d\Shat = \Mhat d\log\Lhat + \log\lambda ds
          +\sum_{n\ge 1} \B_n dz_n +\sum_{n\ge 1} \Bhat_n d\zhat_n.
                                                           \tag 28 \\
\endalign
$$
Actually, $S$ and $\Shat$ mostly play rather technical roles,
however it seems likely that they also have some deep meaning.
Comparing with the Laurent expansion of $\M$ and $\Mhat$, one can
see that $S$ and $\Shat$ have Laurent expansion of the following form.
$$
\align
    & S = \sum_{n\ge 1} z_n\L^n + s\log\L + \sum_{n\le -1} v_n\L^n/n,
                                                            \tag 29 \\
    & \Shat = \sum_{n\ge 1} \zhat_n\Lhat^{-n} + s\log\Lhat + \phi
               + \sum_{n\ge 1} \vhat_n\Lhat^n/n.
                                                            \tag 30 \\
\endalign
$$
This expansion is reminiscent of the Fourier expansion of free bosons
in the KP hierarchy [11]. Remarkably, $\phi$ now arise as a
``zero-mode" and $s$ may be interpreted as its ``conjugate variable."
Finally, the tau function $\tau$ can be defined as
$$
    d\log\tau = \sum_{n\ge 1} v_{-n}dz_n
               -\sum_{n\ge 1} \vhat_n d\zhat_n + \phi ds.   \tag 31
$$

\proclaim{Theorem 1} The right hand side of the above equations
defining $\phi$, $S$, $\Shat$ and $\tau$ are all closed differential
forms.
\endproclaim

One should note here that all the ingredients of the SDiff(2)
Toda hierarchy, $\L$, $\M$, $\Lhat$, $\Mhat$ and $\phi$, can be
reproduced from the tau function.  Indeed, $\phi$ and the Laurent
coefficients $v_n$ and $\vhat_n$ of $\M$ and $\Mhat$ are now given by
a derivative of $\log\tau$, see (31). The Laurent coefficients of
$\L$ and $\Lhat$, as (25) shows, can be written in terms of derivatives
of $\phi$, and $\phi$ is already a derivative of $\log\tau$.
The whole hierarchy can thus, in principle, be rewritten into a
system of nonlinear differential equations with just a single unknown
function, $\tau$. In the case of the KP and TL hierarchy [11] [12],
these equations take the form of Hirota's bilinear equations.
We do not yet know if such a beautiful structure persists in the
SDiff(2) Toda hierarchy.

\heading
    4. Riemann-Hilbert problem and SDiff(2) symmetries
\endheading

\noindent
The nonlinear graviton construction of Penrose [18] can be reformulated
in the present setting as follows.  As basic relation (21) shows,
the two pairs of ``Darboux coordinates" $(\L,\M)$ and $(\Lhat,\Mhat)$
should be connected by a SDiff(2) group element, i.e., an area-preserving
diffeomorphism.  This means that there are two functions $f=f(\lambda,s)$
and $g=g(\lambda,s)$ such that
$$
    f(\L,\M) = \Lhat, \quad  g(\L,\M) = \Mhat                  \tag 32
$$
and
$$
    \{ f, g \} = f.                                            \tag 33
$$
More precisely, we have to impose some analyticity conditions to
$\L$, $\M$, $\Lhat$ and $\Mhat$ so that functional relations (32) are
meaningful. It is customary to assume that the Laurent series giving
$\L$ etc. have a common domain of convergence, say a neighborhood of
a circle $\Gamma$ as mentioned in Section 2. On the complex $\lambda$
plane, then, $\L$ and $\M$ can be analytically extended outside
$\Gamma$ whereas $\Lhat$ and $\Mhat$ inside $\Gamma$.
The ``patching functions'' $f$ and $g$ will then become holomorphic
functions in a neighborhood of $\Gamma$. (We do not
specify the domain of the $s$ plane where they are to be defined;
a rigorous formulation should be given in the language of sheaf
cohomology [19].) Conversely, given a pair of patching functions
satisfying (33), one may ask if functional equations (32)
for $\L$, $\M$, $\Lhat$ and $\Mhat$ have a solution. This is a
kind of Riemann-Hilbert problem, now referring to the SDiff(2) group.
As Penrose [18] first observed, this kind of Riemann-Hilbert problem has
a unique solution as far as the area-preserving diffeomorphism $(f,g)$
is sufficiently close the identity map.

Having this correspondence between a solution of the SDiff(2) Toda
hierarchy and a pair of patching functions, we can apply the previous
method for the self-dual vacuum Einstein equation [14] to the present
setting with slightest modifications.  According to a general prescription
presented therein, we now consider left and right translations caused by
two Hamiltonian vector fields
$$
\align
    & \{F,\cdot\}     = \lambda (\del_\lambda F)\del_s
                       -\lambda (\del_s F) \del_\lambda,             \\
    & \{\Fhat,\cdot\} = \lambda (\del_\lambda \Fhat)\del_s
                      -\lambda (\del_s \Fhat) \del_\lambda   \tag 34 \\
\endalign
$$
as
$$
    (f,g) \longrightarrow
            \exp \epsilon \{\Fhat,\cdot\} \circ (f,g) \circ
            \exp (-\epsilon) \{F,\cdot\},                    \tag 35
$$
where $\epsilon$ is an infinitesimal parameter; ``$\circ$" stands for
composition of diffeomorphisms; $F=F(\lambda,s)$ and
$\Fhat=\Fhat(\lambda,s)$ are arbitrary functions with the same
analyticity as the patching functions.  This should give rise
to an infinitesimal variation of the corresponding solution of the
Riemann-Hilbert problem. From the construction, one can expect a
chiral structure in these symmetries, i.e., left and right components
give rise to two distinct symmetries that commute each other. Expanded
to the first order of $\epsilon$, $\L$, $\M$, $\Lhat$ and $\Mhat$ will
transform as
$$
\align
  & \L \longrightarrow \L + \epsilon\delta\L + O(\epsilon^2), \quad
    \M \longrightarrow \M + \epsilon\delta\M + O(\epsilon^2),   \\
  & \Lhat \longrightarrow \Lhat + \epsilon\delta\Lhat + O(\epsilon^2), \quad
    \Mhat \longrightarrow \Mhat + \epsilon\delta\Mhat + O(\epsilon^2).
                                                              \tag 36  \\
\endalign
$$
The coefficients of $\epsilon$ thus define a linear operator
$\delta=\delta_{F,\Fhat}$ that gives an infinitesimal symmetry of
the SDiff(2) Toda hierarchy.  From a differential-algebraic point of
view [20], $\delta_{F,\Fhat}$ is an operator that acts on $\L$, $\M$,
$\Lhat$ and $\Mhat$ as an abstract derivation. By definition,
this is an inner symmetry,
$$
    \delta_{F,\Fhat}\lambda = \delta_{F,\Fhat}s
    = \delta_{F,\Fhat}z_n = \delta_{F,\Fhat}\zhat_n = 0.     \tag 37
$$
In fact, we have the following very explicit result.

\proclaim{Theorem 2} The infinitesimal symmetries of $\L$, $\M$,
$\Lhat$, $\Mhat$ are give by
$$
\align
  \delta_{F,\Fhat}K
    &= \{ F(\L,\M)_{\le -1}-\Fhat(\Lhat,\Mhat)_{\le -1}, K \} \quad
    \text{for}\ K=\L,\M,                                            \\
  \delta_{F,\Fhat}K
    &= \{ \Fhat(\Lhat,\Mhat)_{\ge 0}-F(\L,\M)_{\ge 0}, K \} \quad
    \text{for}\ K=\Lhat,\Mhat.                             \tag 38  \\
\endalign
$$
\endproclaim

We now show that these infinitesimal symmetries can be extended to
the potentials that we have introduced in the previous section.
A priori, this is by no means obvious, because the potentials
are defined only up to an integration constant, hence one has to show
a way to fix this ambiguity in the course of transformation.  We first
show a result on $\phi$.

\proclaim{Theorem 3} An infinitesimal transformation of $\phi$
consistent with the previous one for $\L$, $\M$, $\Lhat$ and $\Mhat$
is given by
$$
  \delta_{F,\Fhat}\phi
    = -\res F(\L,\M)d\log\lambda +\res \Fhat(\Lhat,\Mhat)d\log\lambda.
                                                            \tag 39
$$
\endproclaim

\noindent
``Consistency" in the statement of the theorem means that the infinitesimal
symmetry retains the basic relation, (25), connecting $\phi$ with other
quantities. This result also shows how $S$ and $\Shat$ should be
transformed. Finally, we have the following result for the tau function.

\proclaim{Theorem 4} An infinitesimal transformation of $\tau$
consistent with the previous one for $\L$, $\M$, $\Lhat$, $\Mhat$
and $\phi$ is given by
$$
    \delta_{F,\Fhat}\log\tau
    = -\res F^s(\L,\M)d_\lambda\log\L
      +\res \Fhat^s(\Lhat,\Mhat)d_\lambda\log\Lhat,        \tag 40
$$
where $d_\lambda$ stands for total differentiation with respect
to $\lambda$, and $F^s=F^s(\lambda,s)$ and $\Fhat^s=\Fhat^s(\lambda,s)$
are given by
$$
    F^s(\lambda,s) \=def \int_0^s F(\lambda,\sigma)d\sigma, \quad
    \Fhat^s(\lambda,s) \=def \int_0^s \Fhat(\lambda,\sigma)d\sigma.
                                                           \tag 41
$$
\endproclaim

Commutation relations for these infinitesimal symmetries can also be
calculated.

\proclaim{Theorem 5} The infinitesimal symmetries $\delta_{F,\Fhat}$
on $\L$, $\M$, $\Lhat$, $\Mhat$ and $\phi$ respect the SDiff(2)
structure in the sense that the commutation relations
$$
    \left[ \delta_{F_1,\Fhat_1}, \delta_{F_2,\Fhat_2} \right]K
    = \delta_{ \{F_1,F_2\}, \{\Fhat_1,\Fhat_2\} }K        \tag 42
$$
are satisfied for $K=\L,\M,\Lhat,\Mhat, \phi$.
\endproclaim

\noindent
This result is not very surprising, because we already know of
a similar result in the case of the self-dual vacuum Einstein
equations [14].  For several reasons, $\phi$ may be considered
an analogue of the Plebanski (first and second) key functions,
and it is shown therein that infinitesimal symmetries of the above
type can be extended to the key functions without any anomaly.
Anomalous commutation relations, however, take place at the level of
the tau function.

\proclaim{Theorem 6} For the tau function,
$$
    \left[ \delta_{F_1,\Fhat_1}, \delta_{F_2,\Fhat_2} \right]\log\tau
    = \delta_{ \{F_1,F_2\}, \{\Fhat_1,\Fhat_2\} }\log\tau
     +c(F_1,F_2) + \chat(\Fhat_1,\Fhat_2),                \tag 43
$$
where $c$ and $\chat$ are cocycles of the SDiff(2) algebra given by
$$
\align
    & c(F_1,F_2) \=def -\res F_2(\lambda,0)dF_1(\lambda,0),     \\
    & \chat(\Fhat_1,\Fhat_2)
         \=def \res \Fhat_2(\lambda,0)d\Fhat_1(\lambda,0). \tag 44
\endalign
$$
\endproclaim

\noindent
We are thus naturally led to a central extension of the SDiff(2)
algebra (or, more precisely, the direct sum of two copies of the
SDiff(2) algebra).  The cocycles are of Kac-Moody type and give
rise to a U(1) current algebra in the ``spin 1" sector of
SDiff(2).  It is amusing to compare the above result with
physicists' calculation of cocycles for SDiff(2) algebras on
various surfaces [21] [22] [23] [24]; they observed that there are exactly
$2g$ linearly independent cocycles on a genus $g$ surface.  Since
the cylinder $S^1 \times \R$ may be thought of as a genus $g=1/2$ surface,
our result seems to fit well into physicists' observation.

\heading
    5. Conclusion
\endheading

\noindent
We have thus introduced the SDiff(2) Toda hierarchy and shown that
it shares a number of remarkable characteristics with the ordinary
KP/TL hierarchies.  In particular, SDiff(2) symmetries of the tau
function exhibit commutator anomalies, hence the tau function requires
a central extension of SDiff(2) as a true symmetry algebra.  On the other
hand, the Lax formalism and related technical details such as the
Riemann-Hilbert problem are obviously of the same type as the
self-dual vacuum Einstein equations and its hyper-K\"ahler version.
We thereby expect that the SDiff(2) Toda equation (hierarchy) will
be a nice laboratory for an attempt to unify two apparently distinct
families of nonlinear ``integrable" systems, soliton equations and
nonlinear graviton equations, on an equal footing.

One can deduce a similar conclusion for Krichever's SDiff(2) version
of the KP hierarchy [17]. This issue will be reported elsewhere.

\rmproclaim{Acknowledgement} One of the authors (K.T.) is very grateful
to J.D. Finley, I. Bakas and Q-Han Park for many valuable suggestions.
This work was completed in the workshop ``Project RIMS 91" held at the
Research Institute for Mathematical Sciences, Kyoto University, during
June -- August 1991.


\heading
    References
\endheading
\baselineskip=16pt

\item{1.}
Saveliev, M.V., and Vershik, A.M.,
Continual analogues of contragredient Lie algebras,
Commun. Math. Phys. 126 (1989), 367-378.
\item{2.}
Bakas, I.,
The structure of the $W_\infty$ algebra,
Commun. Math. Phys. 134 (1990), 487-508.
\item{3.}
Park, Q-Han, Extended conformal symmetries in real heavens,
Phys. Lett. 236B (1990), 429-432.
\item{4.}
Boyer, C., and Finley, J.D.,
Killing vectors in self-dual, Euclidean Einstein spaces,
J. Math. Phys. 23 (1982), 1126-1128.
\item{5.}
Gegenberg, J.D., and Das, A.,
Stationary Riemannian space-times with self-dual curvature,
Gen. Rel. Grav. 16 (1984), 817-829.
\item{6.}
Hitchin, N.J.,
Complex manifolds and Einstein's equations,
in {\it Twistor Geometry and Non-linear Systems\/},
H.D. Doebner and T. Weber (eds.), Lecture Notes in Mathematics  vol. 970
(Springer-Verlag 1982).
\item{7.}
Jones, P.E., and Tod, K.P.,
Minitwistor spaces and Einstein-Weyl spaces,
Class. Quantum Grav. 2 (1985), 565-577.
\item{8.}
Ward, R.S.,
Einstein-Weyl spaces and $SU(\infty)$ Toda fields,
Class. Quantum Grav. 7 (1990). L95-L98.
\item{9.}
LeBrun, C.,
Explicit self-dual metrics on $CP_2$ \# $\dots$ \# $CP_2$,
J. Diff. Geometry (to appear).
\item{10.}
Sato, M., and Sato, Y.,
Soliton equations as dynamical systems in an infinite dimensional
Grassmann manifold,
in {\it Nonlinear Partial Differential Equations in Applied Sciences\/},
P.D. Lax, H. Fujita, and G. Strang eds.
(North-Holland, Amsterdam, and Kinokuniya, Tokyo, 1982).
\item{11.}
Date, E., Kashiwara, M., Jimbo, M., and Miwa, T.,
Transformation groups for soliton equations,
in {\it Nonlinear Integrable Systems --
Classical Theory and Quantum Theory\/},
M. Jimbo and T. Miwa eds.
(World Scientific, Singapore, 1983).
\item{12.}
Ueno, K., and Takasaki, K.,
Toda lattice hierarchy,
in {\it Group Representations and Systems of Differential Equations\/},
Advanced Studies in Pure Mathematics vol. 4
(Kinokuniya, Tokyo, 1984).
\item{13.}
Takebe, T.,
Toda lattice hierarchy and conservation laws,
Commun. Math. Phys. 129 (1990), 281-318.
\item{14.}
Takasaki, K.,
Symmetries of hyper-K\"{a}hler (or Poisson gauge field) hierarchy,
J. Math. Phys. 31 (1990), 1877-1888.
\item{15.}
Kashaev, R.M., Saveliev, M.V., Savelieva, S.A., and Vershik, A.M.,
On nonlinear equations associated with Lie algebras of diffeomorphism
groups of two-dimensional manifolds,
Institute for High Energy Physics preprint 90-I (1990).
\item{16.}
Golenisheva-Kutuzova, M.I., and Reiman, A.G.,
Integrable equations related to Poisson algebras,
Zap. Nauch. Semin. LOMI 169 (1988), 44 (in Russian).
\item{17.}
Krichever, I.M.,
The dispersionless Lax equations and topological minimal models,
preprint (1991).
\item{18.}
Penrose, R.,
Nonlinear gravitons and curved twistor theory,
Gen. Rel. Grav. 7 (1976), 31-52.
\item{19.}
Park, Q-Han,
Self-dual Yang-Mills (+gravity) as a 2D sigma model,
Phys. Lett. B257 (1991), 105-110.
\item{20.}
Takasaki, K.,
Differential algebras and ${\cal D}$-modules
in super Toda lattice hierarchy,
Lett. Math. Phys. 19 (1990), 229-236.
\item{21.}
Arakelyan, T.A., and Savvidy, G.K.,
Cocycles of area-preserving diffeomorphisms and anomalies in the
theory of relativistic surfaces,
Phys. Lett. 214B (1988), 350-356.
\item{22.}
Bars, I., Pope, C.N., and Sezgin, E.,
Central extensions of area preserving membrane algebras,
Phys. Lett. 210B (1988), 85-91.
\item{23.}
Floratos, F.G., and Iliopoulos, J.,
A note on the classical symmetries of the closed bosonic membranes,
Phys. Lett. 201B (1988), 237-240.
\item{24.}
Hoppe, J.,
Diff${}_AT^2$, and the curvature of some infinite dimensional
manifolds,
Phys. Lett. 215B (1988), 706-710.

\bye